\documentclass[a4paper,fleqn]{cas-dc}

\usepackage[authoryear]{natbib}
\usepackage{tikz}
\usepackage{standalone}
\usetikzlibrary{arrows.meta}
\tikzset{cross/.style={cross out, draw, 
         minimum size=2*(#1-\pgflinewidth), 
         inner sep=0pt, outer sep=0pt}}
\usepackage{xurl}

\newcommand{\DBH}{D_\text{BH}}
\begin{document}
\let\WriteBookmarks\relax
\def\floatpagepagefraction{1}
\def\textpagefraction{.001}
% \shorttitle{Wave attenuation in vertically heterogeneous mangroves}
\shorttitle{Root Morphology and Mangrove Wave Attenuation}
\shortauthors{K.E. Pang and Z.Y. Tay}

% \title [mode = title]{On the modelling of the hydrodynamic drag of mangroves} 
% \title [mode = title]{Wave attenuation in vertically heterogeneous mangroves under sea-level rise}
\title [mode = title]{Modelling Wave Attenuation by Mangroves: Effects of Root Morphology and Sea-Level Rise}

% \tnotemark[1,2]
% \tnotetext[1]{This document is the results of the research
%    project funded by the National Science Foundation.}
% \tnotetext[2]{The second title footnote which is a longer text matter
%    to fill through the whole text width and overflow into
%    another line in the footnotes area of the first page.}

\author[1]{Khang Ee {Pang}}[orcid=0000-0003-2003-5369]
\ead{kenny.pang@singaporetech.edu.sg}
\credit{Conceptualization, Methodology, Software, Validation, Formal analysis, Writing - Original Draft, Writing - Review \& Editing, Visualization}

\author[1]{Zhi Yung {Tay}}[]
\cormark[1]
\ead{zhiyung.tay@singaporetech.edu.sg}
\credit{Writing - Review \& Editing, Supervision, Funding acquisition}

\affiliation[1]{
    organization={Engineering Cluster, Singapore Institute of Technology},
    addressline={1 Punggol Coast Road}, 
    postcode={828608}, 
    country={Singapore}
    }

\cortext[cor1]{Corresponding author}

\begin{abstract}
Mangroves are increasingly promoted as nature-based solutions for coastal protection, yet many existing models neglect the vertical variation of vegetation biomass, leading to oversimplified representations of root–flow interactions. In this study, we introduce a generalised parametrisation of the mangrove vegetation profile and derive a wave attenuation model that explicitly accounts for the mangrove root characteristics. The model reveals that wave attenuation is species specific and exhibits a non-monotonic dependence on wave frequency. We further quantify the sensitivity of wave attenuation efficiency to sea-level rise and show that rising sea levels do not necessarily reduce attenuation capacity. In particular, when the water depth is less than approximately $1.6$ times the mean root height, wave attenuation may instead be enhanced under sea-level rise. Future changes in attenuation efficiency under different climate scenarios are also quantified. OpenFOAM simulations are further used to calibrate drag coefficients as a function of the root submergence ratio. These results highlight the critical role of vertical root structure in governing wave attenuation and provide a framework for assessing the long-term resilience of mangrove-based coastal protection under climate change. 
\end{abstract}

% \includegraphics{figs/cas-grabs.pdf}

% \begin{highlights}
% \item Research highlights item 1
% \item Research highlights item 2
% \item Research highlights item 3
% \end{highlights}


\maketitle

\section{Introduction}
\label{sec:intro}

Mangrove forests are increasingly recognised as effective nature-based solutions (NbS) for coastal protection. By extracting momentum and dissipating wave energy through their complex above-ground root systems, mangroves can significantly reduce wave heights before they reach the shoreline \citep{mazda_mangrove_1997}. In addition to mitigating coastal hazards, mangroves provide a range of ecosystem services including carbon sequestration \citep{alongi2014}, habitat provision \citep{Nagelkerken2008}, and shoreline stabilisation \citep{alongi2008}. As sea levels continue to rise, understanding the hydrodynamic performance of mangroves under changing environmental conditions has become increasingly important for coastal adaptation planning.

For surface gravity waves, wave energy is concentrated primarily in the upper portion of the water column, causing wave attenuation to be particularly sensitive to the vertical distribution of vegetation-induced drag \citep{mazda_wave_2006}. Because wave orbital velocities decay with depth, vegetation elements located near the water surface interact more strongly with the flow than those near the bed. Consequently, two mangrove trees with the same frontal area may exhibit substantially different wave attenuation characteristics depending on the vertical arrangement of their root systems. The vegetation-induced drag force per unit volume at elevation $z$ is given by
\begin{equation}
f_D = \frac{1}{2}\rho C_D N_\text{tree} a u|u|,
\end{equation}
where $a(z)$ is the height-resolved vegetation width profile per tree, $C_D$ is the drag coefficient, $N_\text{tree}$ is the number of trees per $\mathrm{m}^2$, $\rho$ is the fluid density, and $u$ is the horizontal fluid velocity.

Assuming a depth-constant vegetation profile, \citet{dalrymple1984} derived an analytical expression for wave decay within a vegetated region by assuming a constant rate of wave energy dissipation per unit volume. \citet{mendez_losada_2004} subsequently extended the formulation to random waves and sloping bathymetry. Owing to its analytical tractability, the Dalrymple model has become a cornerstone of vegetation wave attenuation theory and forms the basis of many subsequent studies.

Recent investigations have highlighted the importance of resolving the vertical structure of vegetation when modelling vegetated flows. Experimental \citep{wu2015} and numerical \citep{wu2016} works demonstrated that depth-uniform vegetation profiles can significantly overestimate wave attenuation compared with vertically heterogeneous vegetation, while \citet{horstman2018} discussed the limitations of homogeneous cylinder arrays with constant height in representing natural mangrove root systems. These findings suggest that accurately characterising the vertical distribution of vegetation is essential for reliable prediction of wave attenuation.

Several parametric models have been proposed to describe the root geometry of \textit{Rhizophora} mangroves \citep{ohira2013,yoshikai2021} and have been adopted in laboratory experiments. However, physical mangrove prototypes often fail to reproduce the observed frontal-area distribution of real vegetation, particularly near the bed \citep{fernando2024}. Furthermore, existing geometric models are typically species-specific and rely on detailed morphological parameterisations that are difficult to incorporate directly into analytical wave attenuation theories. Thus, there remains a need for a simple and generalised representation that captures the essential characteristics of mangrove root morphology while remaining analytically tractable.

The challenge becomes even more relevant under projected sea-level rise. Increasing water depth modifies both the wave kinematics and the degree of vegetation submergence, potentially altering the wave attenuation capacity of mangrove forests. While previous studies have quantified wave attenuation under present-day conditions, comparatively little attention has been given to how mangrove root morphology influences the sensitivity of wave attenuation to sea-level rise. A quantitative framework capable of linking vegetation structure, wave attenuation, and future sea-level-rise scenarios would therefore be valuable for assessing the long-term resilience of mangrove-based coastal protection.

In this study, we introduce a generalised parametrisation of the vegetation profile $a(z)$ that can be calibrated from field observations and applied across different mangrove species, sites, and geographical regions. Using this representation, we derive an analytical wave attenuation model that explicitly accounts for the vertical distribution of mangrove roots and investigate the dependence of wave attenuation on wave regime and vegetation morphology. We further quantify the sensitivity of wave attenuation efficiency to sea-level rise and assess future changes in attenuation capacity using IPCC projections. Finally, we develop an equivalent representation based on vertically distributed cylindrical elements that can be readily implemented in laboratory experiments and numerical simulations, and use this framework to calibrate drag coefficients associated with different vegetation profiles.

\section{Theoretical model}
\label{sec:theory}

\subsection{Vertical mangrove profile}

Considering the recursively branching nature of the mangrove root system, we model the root density as an exponential function such that
\begin{align} \label{eq:w1}
    a(z) &= \underbrace{a_0\mathrm{e}^{-\beta z}}_{\text{root}} + \underbrace{\DBH}_{\text{trunk}}, \qquad z\geq0,
\end{align}
where $\beta$ is the root characteristic parameter that governs the steepness of the vegetation profile and $a_0$ is the total root width at $z=0$ (per tree). We shall see, in Section~\ref{sec:construction}, that $1/\beta $ can be interpreted as the mean root height, and the ratio $\beta h$ is ratio between the water depth to the mean root height. Furthermore, if we know the mean root diameter as a function of elevation $D(z)$, then the number of roots at elevation $z$ is given by
\begin{equation}
    n(z) = \frac{a_0}{D(z)}\mathrm{e}^{-\beta z}, \qquad z\geq0.
\end{equation}
We note that the analysis of the trunk contribution in \eqref{eq:w1} is identical to that of \citep{dalrymple1984,mendez_losada_2004}, as such, our analysis shall focus mainly on the root contribution. 

\begin{figure}
	\centering
	\includegraphics[width=.9\columnwidth]{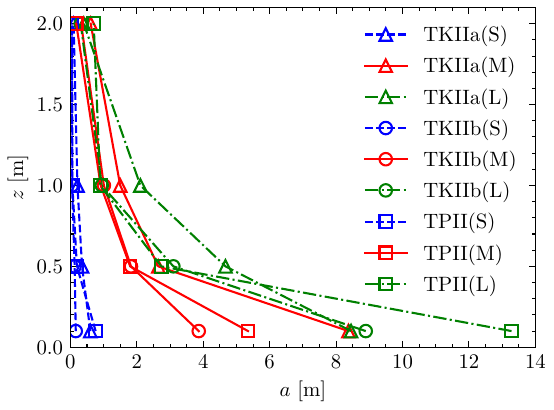}
	\caption{Vegetation profile of \textit{Rhizophora} trees at various sites (TKIIa, TKIIb, TPII) and of various sizes (S, M, L). We refer the reader to \citet{horstman2014} for more details.}
        \label{fig:horstman_semilog}
\end{figure}

\begin{figure}
	\centering
	\includegraphics[width=0.9\columnwidth]{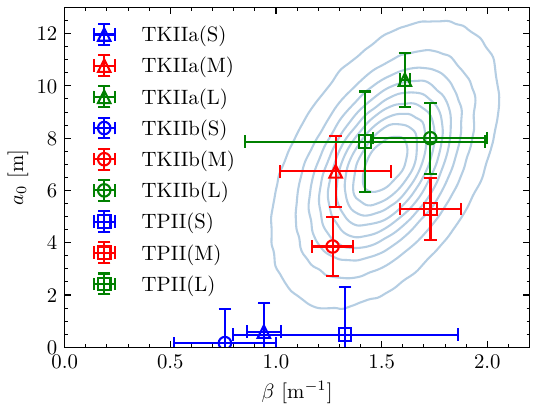}
    \caption{Optimal parameters ($\beta$ and $a_0$) fitted to each observation. A bivariate normal distribution is fitted to the mature (M and L) observations.}
    \label{fig:parameter_fitting}
\end{figure}

Here, we justify the choice of \eqref{eq:w1} using field measurements of \textit{Rhizophora} (stilt roots) and \textit{Sonneratia} (pneumatophores) vegetation profile. Figure~\ref{fig:horstman_semilog} presents the vegetation profiles of nine individual \textit{Rhizophora} trees reported by \citet{horstman2014}. The trees were grouped into three size classes - small (S), medium (M), and large (L) - and sampled from different locations within the Kantang (TKII) and Palian (TPII) estuaries on the southwestern coast of Thailand. 
% $a(z)$ is calculated by multiplying the number of roots by the average root diameter at various elevations ($z=0.1, 0.5,1.0,2.0\,\mathrm{m}$). We observed a monotonic decrease of the frontal width as $z$ increases, and the profile is far from being a constant. 
Next, we fit the observations to our model \eqref{eq:w1} (taking $\DBH=0$) to obtain the optimal parameters and the standard error associated to the parameters. The result is presented in Figure \ref{fig:parameter_fitting}. We find that the variation of $\beta$ is larger for the saplings (S) compared to the mature (M and L) trees. As such, we fit a bivariate normal distribution only to the mature observations (overlaid in Figure \ref{fig:parameter_fitting}) to obtain our model parameters for \textit{Rhizophora}. 
%Figure \ref{fig:parity} compares the measured vegetation profiles to our model predictions.
Similar analysis were performed for \textit{Sonneratia} pneumatophore using data from \citet{lienard2016}. The recommended parameters are summarised in Table \ref{tab:recommendation}.

\begin{table}[htb]
    \centering
    \begin{tabular}{c c c c c}
        \hline 
        & \multicolumn{2}{c}{\textit{Rhizophora}} & \multicolumn{2}{c}{\begin{tabular}{c}\textit{Sonneratia} \\(pneumatophore)\end{tabular}} \\
        & $a_0\,[\mathrm{m}]$ & $\beta\,[\mathrm{m}^{-1}]$ & $a_0N_\text{tree}\,[\mathrm{m}^{-1}]$ & $\beta\,[\mathrm{m}^{-1}]$ \\
        \hline
        Mean & 7 & 1.5 & 0.9 & 9 \\
        Std & 2 & 0.2 & 0.4 & 2 \\
        \hline
    \end{tabular}
    \caption{Recommended model parameters.}
    \label{tab:recommendation}
\end{table}

\begin{figure}
    \centering
    \includegraphics[width=0.7\columnwidth]{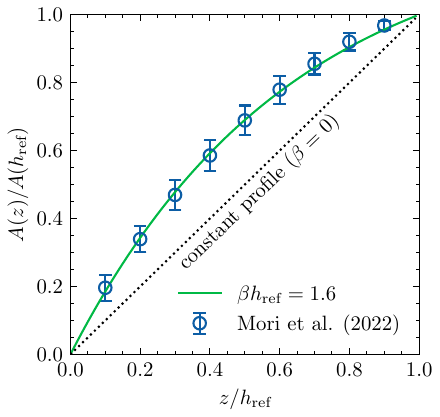}
    \caption{Normalised frontal area of \textit{Rhizophora} as a function of elevation. } % We refer the reader to \cite{mori2022} for more details.}
    \label{fig:mori}
\end{figure}

Meanwhile, we can compute the submerged frontal area as 
\begin{equation}
    A(h) = \int_0^h a(z)\,\mathrm{d}z = \frac{a_0}{\beta }(1-\mathrm{e}^{-\beta h}) + h\DBH.
\end{equation}
Let $h_\text{ref}$ be the height of the tallest root (again ignoring $\DBH$ term), then the normalised submerged frontal area has the form
\begin{equation} \label{eq:norm-ref-area}
    \frac{A(z)}{A(h_\text{ref})} = \frac{1-\mathrm{e}^{-\beta h_\text{ref}z/h_\text{ref}}}{1-e^{-\beta h_\text{ref}}}, \qquad 0\leq z/h_\text{ref}\leq1,
\end{equation}
which eliminates the dependency on $a_0$. Figure \ref{fig:mori} compares \eqref{eq:norm-ref-area} to field measurements by \citet{mori2022}. In particular, we found that $\beta h_\text{ref}=1.6$ is \textit{universal} for \textit{Rhizophora} of all sizes. % In comparison, the model by Ohira et al. underestimates the reference area at a lower elevation due to only considering the primary roots. 
These comparisons give us confidence that \eqref{eq:w1} is a good description of the vegetation profile of mangroves. 

\subsection{Wave attenuation model}

Following \cite{dalrymple1984}, we assume zero current and a constant rate of wave energy dissipation through the vegetation zone, 
\begin{equation} \label{eq:energy}
    \frac{\mathrm{d}}{\mathrm{d} x}(Ec_g) = -\epsilon, \qquad E=\frac{1}{8}\rho g H(x)^2,
\end{equation}
where $E$ is the wave energy per unit area, $\epsilon$ is the rate of dissipation, $c_g$ is the wave group velocity, $g$ is the gravitational acceleration, and $H(x)$ is the wave height along the vegetated channel. From Airy wave theory, the horizontal orbital velocity is given by % and the surface elevation are given by 
\begin{align}
    u(x,z,t) &= \frac{H}{2}\sigma \frac{\cosh(kz)}{\sinh(kh)}\cos(kx-\sigma t).%, \\ 
    % \eta(x,t) &= \frac{H}{2}\cos(kx-\sigma t). 
\end{align}
Here, $\sigma$ is the wave angular frequency, $k$ is the wave number, $h$ is the water depth. On the other hand, the energy dissipation rate is equivalent to the (time-averaged) rate of work done by the vegetation, 
\begin{align}
    \epsilon &= \frac{1}{T}\int_0^T \int_0^h f_D u\,\mathrm{d}z\,\mathrm{d}t, \\
    &= \frac{2}{3\pi}\rho C_D N_\text{tree}\frac{H^3\sigma^3}{8\sinh^3(kh)}\int_0^ha(z)\cosh^3(kz)\,\mathrm{d}z.
\end{align}
Equation \eqref{eq:energy} now becomes an ordinary differential equation of the form
\begin{equation} \label{eq:regular}
    \frac{\mathrm{d}}{\mathrm{d} x}H = -K_D H^2,
\end{equation}
where where $K_D$ is the wave decay coefficient given by
\begin{align} \label{eq:mykd}
    % K_D &= \frac{1}{3\pi}C_DN_\text{tree}\frac{\sigma^3}{gc_g\sinh^3(kh)}\phi, \\
    K_D &= \frac{2}{3\pi}C_D N_\text{tree}\frac{k^2\tanh^2(kh)}{kh +\tanh(kh)-kh\tanh^2(kh)}\phi, \\
    \phi 
    % &= \frac{1}{\sinh^3(kh)}\int_0^ha(z)\cosh^3(kz)\,\mathrm{d}z. \\
    &= \frac{h}{\sinh^3(kh)}\int_0^1 a(h\tilde{z})\cosh^3(kh\tilde{z})\,\mathrm{d}\tilde{z}.
\end{align}
So far, we have kept $a(z)$ arbitrary. Finally, solving \eqref{eq:regular} for $H(x)$ gives 
\begin{equation} \label{eq:a}
    \frac{H(x)}{H_0} = \frac{1}{1+K_DH_0x}, \qquad x\geq0,
\end{equation}
with $H_0=H(0)$ being the incident wave height.

Previous vegetation models \citep{dalrymple1984,mendez_losada_2004,suzuki2012} typically approximate the vertical vegetation profile $a(z)$ using a small number of layers with piecewise-constant profile. While computationally convenient, such representations introduce additional fitting parameters such as the number of layers and the layer height, and may obscure species-specific morphological effects. The continuous profile \eqref{eq:w1} avoids this discretisation and permits analytical evaluation of the wave attenuation coefficient, which gives
\begin{multline}
    % K_D &= \frac{1}{3\pi}C_D N_\text{tree} \frac{1}{\sinh^3(kh)}\frac{2k^2\tanh^2(kh)}{kh +\tanh(kh)-kh\tanh^2(kh)}\phi, \\ 
    \phi = \frac{a_0h}{8\sinh^3(kh)}\biggl(\frac{\mathrm{e}^{3kh-\beta h}-1}{3kh-\beta h} + 3\frac{\mathrm{e}^{kh-\beta h}-1}{kh-\beta h} \\ +3\frac{\mathrm{e}^{-kh-\beta h}-1}{-kh-\beta h} + \frac{\mathrm{e}^{-3kh-\beta h}-1}{-3kh-\beta h}\biggl).
\end{multline}
% \end{subequations}

In shallow water, the fluid velocity is approximately constant in depth, %with $u=\tfrac{H}{2}\sigma\cos(\sigma t)/kh$ and $c_g=\sqrt{gh}$. 
and the wave decay coefficient reduces to
\begin{equation} \label{eq:w1bar}
    \qquad K_D \sim \frac{1}{3\pi}C_DN_\text{tree}\bar{a}\frac{1}{h}, \qquad kh\ll1,
\end{equation}
where $\bar{a}=A(h)/h$ is the submerged depth-averaged profile. Thus, in the shallow-water regime, $K_D$ is independent of the specific vegetation profile $a(z)$ and only on the average $\bar{a}$. Motivated by this, we introduce the normalised wave decay coefficient as 
\begin{equation} \label{eq:normkd}
    \tilde{K}_D = \frac{3\pi h}{C_DN_\text{tree}\bar{a}}K_D.
\end{equation}
The normalised coefficient $\tilde{K}_D$ only depends on two dimensionless numbers $\beta h$ and $kh$. Such a non-dimensional formulation facilitates the comparison of nature-based systems with different amounts of submerged vegetation. Equation \eqref{eq:normkd} does implicitly assumes a constant drag coefficient $C_D$, although in reality $C_D$ may vary depending on the nature of the flow-vegetation interaction. Nevertheless, investigating this idealised scenario provides valuable insight into the fundamental mechanisms governing wave attenuation by vegetation. %The calibration and variability of $C_D$ will be discussed in a subsequent section.

\begin{figure}
    \centering
    \includegraphics[width=0.9\columnwidth]{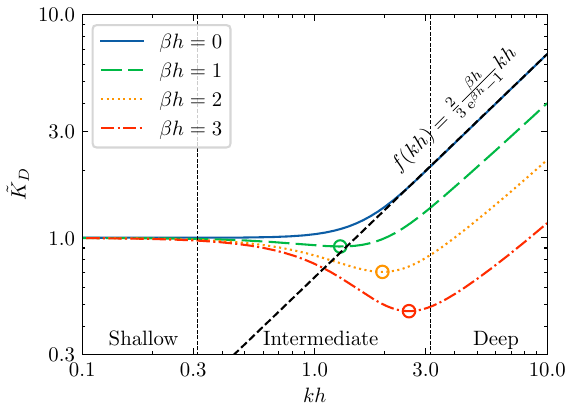}
    \caption{Normalised wave decay coefficient as function of $kh$. Hollow circle marks the global minimum. Both $x$ and $y$ axes are plotted on a logarithmic scale. $C_D$ is taken to be unity.}
    \label{fig:kd}
\end{figure}

Figure \ref{fig:kd} plots \eqref{eq:normkd} using the exponential vegetation profile \eqref{eq:w1} with different values of $\beta h$ on a log-log scale. In the shallow-water regime, $\tilde{K}_D$ is insensitive to $\beta h$ since the wave energy is distributed evenly in depth. In contrast
% , in deep water regime we have 
% \begin{equation}
%     K_D = \frac{4}{3\pi}C_DN_\text{tree} k^2a_0h\frac{\mathrm{e}^{-3kh}-\mathrm{e}^{-\beta h}}{{\beta h-3kh}}, \qquad kh\gg1.
% \end{equation}
% In particular 
for $kh$ dominant cases, we have
\begin{equation} \label{eq:kd_deep}
    K_D\sim\frac{2}{9\pi}C_DN_\text{tree}ka_0\mathrm{e}^{-\beta h}, \qquad kh\gg\beta h, 
\end{equation}
and $\tilde{K}_D$ scales linearly with $kh$ as shown in the black dashed line in Figure \ref{fig:kd}. In relation to the vegetation distribution, increasing $\beta h$ leads to reduced wave attenuation due to fewer of the root engaging with the waves. This trend is consistent with previous works \citep{wu2015,wu2016}. In the intermediate-water regime, our model reveals a ($\beta h$)-dependent critical $kh$ value over which wave attenuation is least effective, marked by the hollow circle in Figure \ref{fig:kd}. This is in contrast to the classical vegetation model (equivalent to $\beta h = 0$), where $\tilde{K}_D$ increases monotonically with $kh$ and is independent of species. 

We remark that the present analysis may also be extended to random wave. Assuming a narrow-banded wave spectrum, \cite{mendez_losada_2004} showed that the evolution of the root-mean-square wave height $H_{\mathrm{rms}}$ is governed by
\begin{equation}
    \frac{\mathrm{d}}{\mathrm{d} x}H_\text{rms} = -\frac{3\sqrt{\pi}}{4}K_D H_\text{rms}^2 \approx -1.329 K_D H_\text{rms}^2,
\end{equation}
where $K_D$ is evaluated using the peak wave number $k_p$ instead of $k$. Relative to the regular-wave result \eqref{eq:regular}, the predicted attenuation rate is approximately $32.9\%$ larger, assuming that the drag coefficient $C_D$ remains unchanged between the regular and random wave cases. 
Regardless, the specific value of $C_D$ is not important for the next piece of analysis. 
% Nevertheless, as will be shown in the next section, the sea-level-rise sensitivity analysis is relatively insensitive to the specific value of $C_D$. 

\subsection{Impact of sea-level rise on mangrove wave attenuation efficiency}
\label{sec:slr}

We now consider a sea-level rise (SLR) scenario where the water depth rises by an amount of $\Delta h$ from the baseline water depth of $h$. We define the wave attenuation efficiency as the ratio of the wave decay coefficient before and after SLR: 
\begin{eqnarray}
    \eta(\Delta h) = \frac{\tilde{K}_D\left(\beta (h+\Delta h), k(h+\Delta h)\right)}{\tilde{K}_D(\beta h, kh)}.
\end{eqnarray}
For short term sea-level rise projection, we can approximate $\eta(\Delta h)$ with a linear Maclaurin expansion 
\begin{equation}
    \eta = 1 + m\frac{\Delta h}{h} + O(\Delta h^2).
\end{equation}
where $m = h\eta'(0)$ characterizes the sensitivity of wave attenuation efficiency to SLR. The explicit expression of $m$ is given in Appendix~\ref{sec:slr_appendix}. Since $\Delta h/h$ is the fractional increase in water depth, $m$ may be interpreted as the percentage change in attenuation efficiency associated with a one-percent increase in water depth. Thus, $m<0$ corresponds to a \textit{reduction} in efficiency under SLR, whereas $m>0$ corresponds to an \textit{improvement} in efficiency. We note that $m$ is dimensionless and depends only on the dimensionless parameters $kh$ and $\beta h$. 

\begin{figure}
    \centering
    \includegraphics[width=0.9\columnwidth]{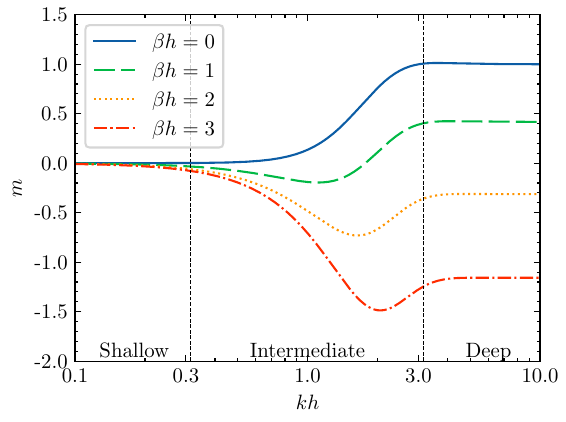}
    \caption{Dimensionless SLR sensitivity $m$ as a function of $kh$ for different values of $\beta h$. The $x$-axis is plotted on a logarithmic scale.}
    \label{fig:slr}
\end{figure}

Figure~\ref{fig:slr} shows $m$ as a function of $kh$ for several values of $\beta h$ on a semi-logarithmic scale. The plot indicate that mangrove wave attenuation is relatively insensitive to SLR in the shallow-water regime due to the weak dependence $\tilde{K}_D$ on both $kh$ and $\beta h$. In contrast, in the deep-water limit, using \eqref{eq:kd_deep} yields
\begin{eqnarray}\label{eq:slr_deep}
    m \sim \frac{2\mathrm{e}^{\beta h}-\beta h\mathrm{e}^{\beta h} -2}{\mathrm{e}^{\beta h}-1}, \qquad kh\gg\beta h.
\end{eqnarray}
This asymptotic expression provides the limiting SLR sensitivity for large $kh$, although it does not necessarily correspond to the extremum across $kh$. 

Moreover, when $\beta h=0$, the attenuation efficiency increases under SLR ($m>0$) for all values of $kh$. Physically, this occurs because the increase in water depth also increases $kh$, leading to stronger attenuation, as seen previously in Figure \ref{fig:kd}. As $\beta h$ increases, this beneficial effect is progressively offset by the increased submergence of the vegetation. Above the critical threshold $\beta h \gtrsim 1.5936$, attenuation efficiency decreases ($m<0$) for all $kh$. The critical value corresponds to the root of \eqref{eq:slr_deep} for which the deep-water sensitivity changes sign. Since $\beta h$ represents the ratio of water depth to mean root height, this threshold suggests that restoration or conservation efforts aimed at maximizing wave attenuation may be most effective in regions where the typical water depth remains below this threshold. 

\begin{figure}
    \centering
    \includegraphics[width=0.9\columnwidth]{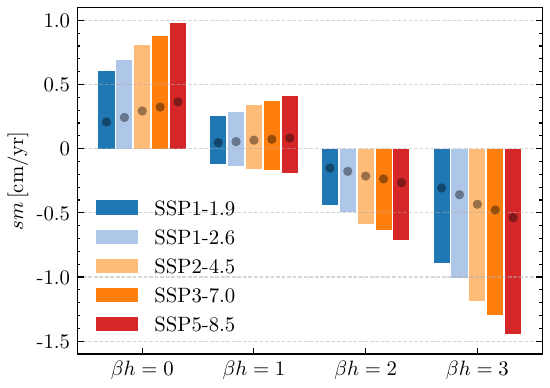}
    \caption{Rate of change of attenuation efficiency based on IPCC AR6 sea-level-rise scenarios. Circle marks the midrange of $m$ with median SLR rate.}
    \label{fig:scenario}
\end{figure}

The sensitivity parameter $m$ also enables straightforward projections of future attenuation efficiency. Let $s={\Delta h}/{\Delta t}$ denote the rate of sea-level rise, then the future mangrove wave attenuation efficiency can be computed as 
\begin{eqnarray}
    \eta = 1 + \frac{sm}{h}\Delta t + O(s^2\Delta t^2).
\end{eqnarray}
% Consequently, ecosystems characterized by larger values of $m$ experience a faster deterioration of wave attenuation capacity under the same SLR scenario. 
Figure~\ref{fig:scenario} presents the 83\% confidence interval of the rate of change of the attenuation efficiency based on the sea-level-rise projections reported in the IPCC Sixth Assessment Report (AR6) \citep[Section 9.6]{fox2023}. As an example, for $h=1\,\mathrm{m}$ and $\beta = 2\,\mathrm{m}^{-1}$, the wave attenuation efficiency loss per annum is between $0.4\%$ to $0.7\%$ for SSP1-1.9 and SSP5-8.5 respectively. The annual SLR rates used in the analysis are summarized in Table \ref{tab:slr}, where the lower and upper confidence of the SLR rates correspond to the 17th and 83rd percentile range respectively. 

\begin{table}[htb] 
    \centering 
    \caption{Projected rates of sea-level rise obtained from the IPCC Sixth Assessment Report \citep{fox2023}.} 
    \begin{tabular}{cccc} 
    \hline 
    Scenario & Lower & Median & Upper \\ 
    & [$\mathrm{cm}/\mathrm{yr}$] & [$\mathrm{cm}/\mathrm{yr}$] & [$\mathrm{cm}/\mathrm{yr}$] \\ 
    \hline 
    SSP1-1.9 & 0.28 & 0.41 & 0.60 \\ 
    SSP1-2.6 & 0.35 & 0.48 & 0.68 \\ 
    SSP2-4.5 & 0.44 & 0.58 & 0.80 \\ 
    SSP3-7.0 & 0.50 & 0.64 & 0.87 \\ 
    SSP5-8.5 & 0.56 & 0.72 & 0.97 \\ 
    \hline 
    \end{tabular} 
    \label{tab:slr} 
\end{table}

Finally, we note that the present analysis assumes that the drag coefficient remains approximately unchanged over the small depth increments associated with short-term SLR. Under this assumption, the ratio defining $\eta$ becomes insensitive to $C_D$, and the leading-order sensitivity $m$ is determined primarily by the dimensionless parameters $kh$ and $\beta h$. 

% \subsection{Can mangrove growth outpace the impact of sea-level rise?}

% We can model the size of the mangrove tree as a function of age using
% \begin{eqnarray}
%     a_0 = a_{\infty}(1-\mathrm{e}^{-\lambda t}), \qquad t\geq0.
% \end{eqnarray}
% where $a_{\infty}$ is the asymptote size of the mangrove when $t\rightarrow\infty$ and $\lambda$ is the growth rate. 

% For the implementation of the mangrove forest, it is important to know weather how fast one could expect the wave attenuation effect of mangrove will take effect from the time of planting. In such cases, one would consider relatively young mangrove forest. In this case it is sufficient to consider the linear approximation
% \begin{equation}
%     a_0(t) \sim a_\infty\lambda t, \qquad \lambda t\ll1. 
% \end{equation}

% The efficiency 
% \begin{eqnarray}
%     \eta(\Delta t) = \frac{K_D(t+\Delta t)}{K_D(t)}=\frac{G(t+\Delta t)}{G(t)}
% \end{eqnarray}

\section{Numerical model}
\label{sec:numerical}

\subsection{A simplified mangrove representation}
\label{sec:construction}

Consider a set of $N_\text{cyl}$ cylinders (per $\mathrm{m}^2$) with constant diameter $D$ and base located at $z=0$. Suppose the height of the cylinders follow an exponential distribution $Z\sim\operatorname{Exp}(\beta )$, then the probability of each cylinder intersecting the $xy$-plane at elevation $z$ is $\mathbb{P}(Z\geq z)$. Using this, the vegetation profile (per $\mathrm{m}^2$) can be calculated as
\begin{align}
    a(z)N_\text{tree} &= DN_\text{cyl}\mathbb{P}(Z\geq z), \\
    &= DN_\text{cyl}\int_z^\infty \beta \mathrm{e}^{-\beta  \tilde{z}}\,\mathrm{d}\tilde{z}, \\
    &= DN_\text{cyl}\mathrm{e}^{-\beta z}. 
\end{align}
Thus, by using only vertical cylindrical rods, we can recover the desired hydrodynamic drag profile \eqref{eq:w1} with $DN_\text{cyl}=a_0N_\text{tree}$. Furthermore, the mean root (cylinder) height is given by $\mathbb{E}[Z]=1/\beta $ and variance $\operatorname{Var}[Z]=1/\beta^2$. The simplicity of such geometry could greatly reduce the fabrication cost of constructing mangrove prototypes. The rest of this section served to justify the proposed design. 

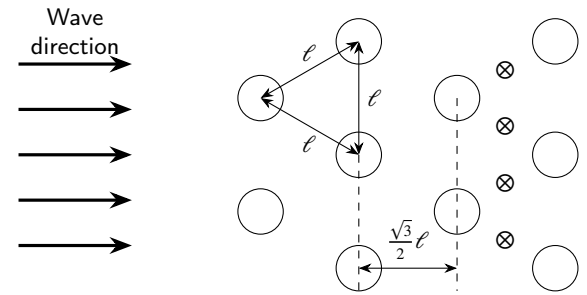
\begin{figure}
    \centering
    \begin{tikzpicture}[scale=1.5]
        \draw (0,0) circle (0.2);
        \draw (0, 1) circle (0.2);
        \draw (0, -1) circle (0.2);
        \draw ({cos(30)},{sin(30)}) circle (0.2);
        \draw ({cos(-30)},{sin(-30)}) circle (0.2) ;
        \draw ({cos(150)},{sin(150)}) circle (0.2);
        \draw ({cos(-150)},{sin(-150)}) circle (0.2);
        \draw (1.732,0) circle (0.2);
        \draw (1.732, 1) circle (0.2);
        \draw (1.732, -1) circle (0.2);
        \draw (1.29, 0.25) node{$\otimes$};
        \draw (1.29, 0.75) node{$\otimes$};
        \draw (1.29, -0.25) node{$\otimes$};
        \draw (1.29, -0.75) node{$\otimes$};
        \draw[dashed] ({cos(30)},{sin(30)}) -- ({cos(30)},-1.2);
        \draw[dashed] (0,0) -- (0,-1.2);
        \draw[Stealth-Stealth] (0,-1) -- (0.866,-1) node[midway,anchor=south]{$\frac{\sqrt{3}}{2}\ell$};
        \draw[Stealth-Stealth](0,1) -- (0,0) node[midway,anchor=west]{$\ell$};
        \draw[Stealth-Stealth]({cos(150)},{sin(150)}) -- (0,0) node[midway,anchor=north]{$\ell$};
        \draw[Stealth-Stealth]({cos(150)},{sin(150)}) -- (0,1) node[midway,anchor=south]{$\ell$};
        \draw[-Stealth,very thick](-3,0) -- (-2,0);
        \draw[-Stealth,very thick](-3,0.4) -- (-2,0.4);
        \draw[-Stealth,very thick](-3,-0.4) -- (-2,-0.4);
        \draw[-Stealth,very thick](-3,0.8) -- (-2,0.8) node[midway,text width=1.6cm,align=center,anchor=south]{Wave\\direction};
        \draw[-Stealth,very thick](-3,-0.8) -- (-2,-0.8);
    \end{tikzpicture}
    \caption{Staggered cylinder configuration used in present work.}
    \label{fig:array}
\end{figure}

Before that, we have to decide how the cylinders are arranged in the plane. We note that different cylinder arrangement would lead to a slightly different drag coefficient $C_D$, e.g., a random arrangement generally has a higher $C_D$ compared to a more orderly arrangement \citep{maza2015}. For the purpose of our simulation, we consider a staggered array of cylinders in a hexagonal arrangement (see Figure \ref{fig:array}) with spacing $\ell$ between cylinders, giving a cylinder density of $N_\text{cyl} = 2/(\ell^2\sqrt{3})$ per $\mathrm{m}^2$. The surface elevation at $x$ is measured by averaging over wave probes across the channel width (represented by the $\otimes$ symbols in Figure \ref{fig:array}). % Here, we have set $\DBH=0$ for simplicity sake, but one can easily introduce a tall cylinder of diameter $\DBH$ with density $N_\text{tree}$ to include the trunk effect.

\subsection{Simulation setup}
\label{sec:setup}

\begin{figure*}
	\centering
    \includestandalone[width=\textwidth]{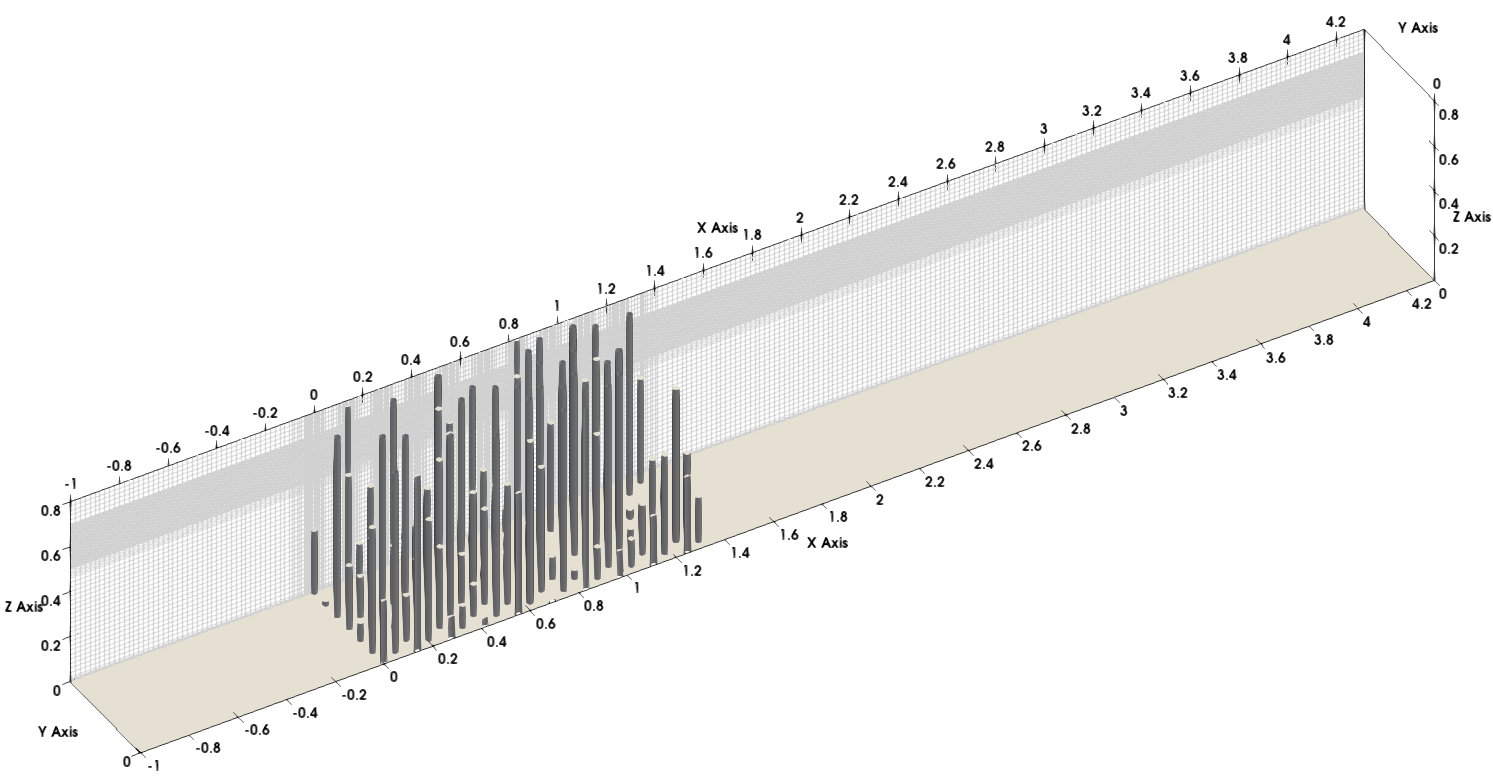}
    \caption{Example mesh used in CFD simulation.}
    \label{fig:mesh}
\end{figure*}

We simulate the wave attenuation in a mangrove forest by solving the unsteady Reynolds-Averaged Navier-Stokes (URANS) equations using the $k$-$\omega$ SST (shear stress transport) turbulence closure model. A multiphase approach is employed via the Volume of Fluid (VOF) method to capture the air-water interface.

Simulations are carried out in OpenFOAM v2412 using the \texttt{interFoam} solver. 
% The \texttt{interFoam} settings suggested in \cite{larsen2019} is used to ensure sufficient convergence of our solution. 
The computational domain is a rectangular cuboid, representing a section of a wave flume or coastal area. The mangrove forest is idealised as an array of vertical rigid cylinders, where the height of the cylinders are generated using the \texttt{numpy.random.exponential} function in Python. No-slip boundary conditions are applied to the bottom and to the cylinder surfaces, atmospheric boundary at the top, and periodic boundaries on the sides. Wave generation and active wave absorption are applied to the inlet and outlet boundaries respectively, based on the method by \citet{higuera2013}. To suppress spurious turbulence generated near the free surface, the turbulence stabilisation method proposed by \citet{larsen2018} is used. 

To ensure high-quality meshing, we use a custom Python script together with the \texttt{blockMesh} utility to generate a structured mesh that conforms to the cylinder array geometry. The initial background mesh has a uniform resolution of $20\,\mathrm{mm}$ in all directions, which is then progressively refined up to two refinement levels near the cylinders, the gas-liquid interface, and the bottom boundary to a resolution of $5\,\mathrm{mm}$. 
%The initial background mesh has a coarsest resolution of $(\Delta x, \Delta y, \Delta z)_\text{coarsest} = (10, 20, 20)\,\mathrm{mm}$, which is progressively refined near the cylinders, the free surface, and the bottom boundary to a finest resolution of $(\Delta x, \Delta y, \Delta z)_\text{finest} = 5\,\mathrm{mm}$. 
Additional boundary layers are constructed around the cylinders to further reduce the first-cell thickness down to $1.8\,\mathrm{mm}$. An example of the generated mesh is shown in Figure~\ref{fig:mesh}.

For the rest of this section, we choose $\beta=1.7\,\mathrm{m}^{-1}$, $a_0=9\,\mathrm{m}$, $D=0.025\,\mathrm{m}$, and $N_\text{tree}=0.5$ trees per $\mathrm{m}^2$ to represent a healthy \textit{Rhizophora}-dominated mangrove forest, giving $\ell=0.08\,\mathrm{m}$ and $N_\text{cyl}=180$ cylinders per $\mathrm{m}^2$.

\subsection{Validation and mesh convergence}

\begin{figure}
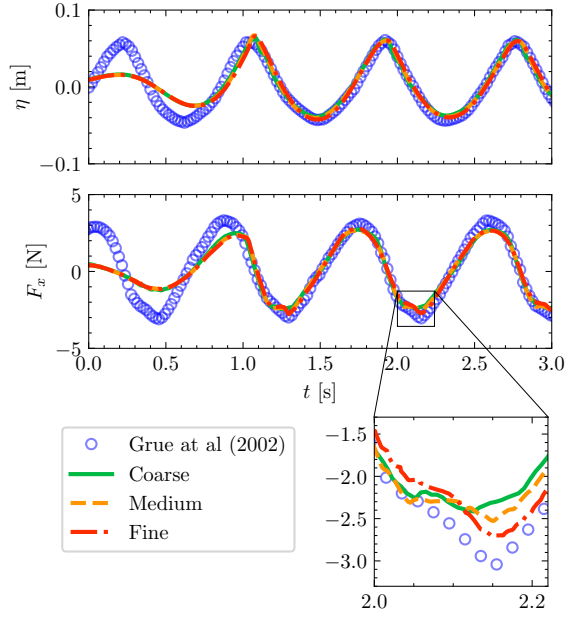

    \centering
    \includestandalone[width=0.9\columnwidth]{tikzpicture/grue}
    \caption{Surface elevation and horizontal wave load on a cylindrical pile. Standard-deviation-normalised RMSE values are given in text.}
    \label{fig:grue}
\end{figure}

We numerically simulate the surface wave load on a single emergent cylinder with diameter $D=0.06\,\mathrm{m}$ under periodic wave conditions $T=0.86\,\mathrm{s}$, $H_0=0.12\,\mathrm{m}$, and $h=0.6\,\mathrm{m}$, matching the experimental setup of \citet{grue2002}. To assess mesh sensitivity, we perform a convergence study using three meshes with finest resolutions of $20\, \mathrm{mm}$ (Coarse), $10\, \mathrm{mm}$ (Medium), and $5\, \mathrm{mm}$ (Fine) in all directions. Figure~\ref{fig:grue} shows good agreement between numerical and experimental results for surface elevation (Normalised RMSE: 0.199, 0.156, 0.125 for Coarse, Medium, Fine) and drag force (Normalised RMSE: 0.179, 0.178, 0.147), confirming sufficient convergence. A similar study is conducted for a cylinder of diameter $D=0.025\,\mathrm{m}$ (not shown here) to further validate convergence. For all subsequent simulations, we adopt the finest mesh resolution of $5\, \mathrm{mm}$.

\subsection{Effect of root shape}

Previous models of Rhizophora roots \citep{ohira2013,yoshikai2021} represent the roots as curved cylinders connected either to neighbouring roots or the tree trunk. In contrast, our model simplifies the geometry by representing the root system using only vertical cylinders. This subsection investigates the influence of root shape on wave attenuation. 

\begin{figure}
    \centering
    \includegraphics[width=1.0\columnwidth]{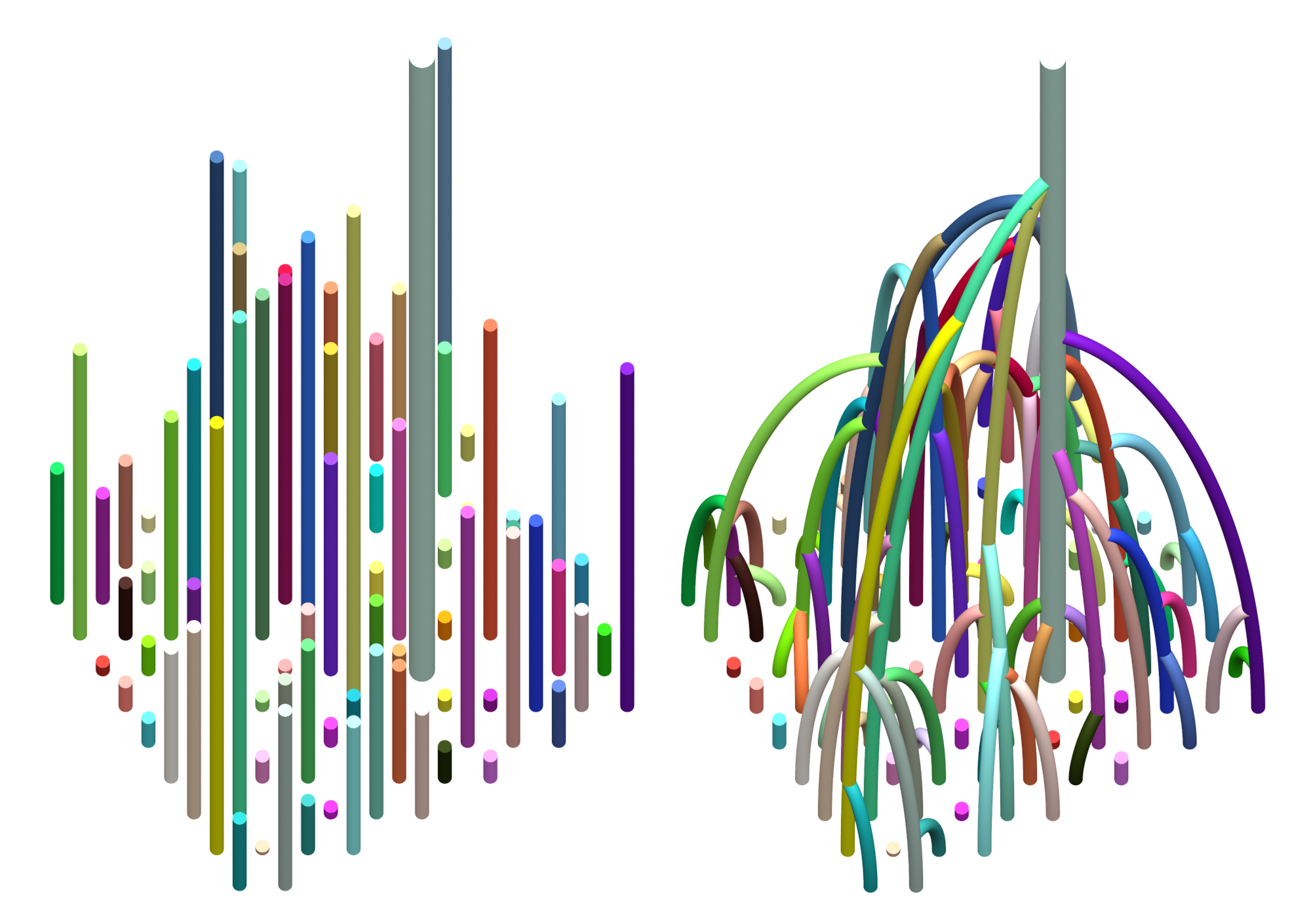}
    \caption{Generated \textit{Rhizophora} tree geometries: straight-root model (left) and curved-root model (right). }
    \label{fig:merge}
\end{figure}

Using the method described in Section \ref{sec:construction}, vertical cylinders are generated within a $0.8\,\mathrm{m}\times0.8\,\mathrm{m}$ area, giving a total of 114 cylinders. The diameter of the tallest root is set to $\DBH=0.046\,\mathrm{m}$ to represent the trunk. To produce the curved roots, each cylinder is bent along an elliptical arc to connect to its nearest taller root. Cylinders shorter than $0.1\,\mathrm{m}$ are excluded from the bending process to avoid generating poor-quality mesh geometry. The resulting geometry before and after the bending operation is shown in Figure~\ref{fig:merge}. The simulation mesh is generated using the \texttt{snappyHexMesh} utility in OpenFOAM with the same two levels of refinement as in the baseline case. 

\begin{figure}
    \centering
    \includegraphics[width=0.8\columnwidth]{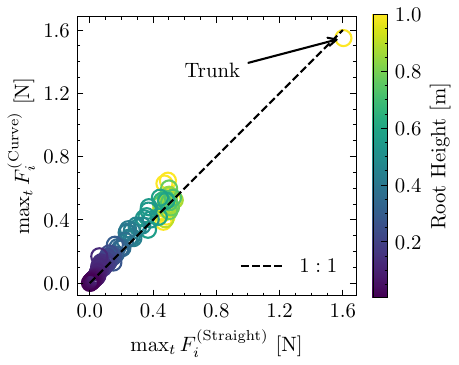}
    \caption{Parity plot comparing the maximum drag force on each root, colour coded by the root height.}
    \label{fig:force_comparison}
\end{figure}

The wave conditions are $T=1.26\,\mathrm{s}$, $H_0=0.12\,\mathrm{m}$, $h=0.6\,\mathrm{m}$, corresponding to $\beta h=1$ and $kh=1.64$. The wave force on each root $F_i(t),\,i=1,\dots,114$, is recorded, and the maximum force of each straight root is compared with that of its curved counterpart in Figure~\ref{fig:force_comparison}. The results show that the total drag force in the curved-root model is only 8.6\% higher than in the simplified straight-root model due to the increased length of the curved roots, suggesting that the straight-root model is a reasonable proxy for mangrove tree. 

% \begin{figure}[htb]
%     \centering
%     \begin{tikzpicture}
%         \node{\includegraphics[width=0.5\textwidth]{images/merging_transparent.png}};
%         \draw[->,thick](-1,1.2) -- (1,1.2) node[midway,anchor=south]{Merge};
%         \draw[<-,thick](-1,1.0) -- (1,1.0) node[midway,anchor=north]{Simplify};
%         % \draw (-3.2,2) node{P1};
%         % \draw (3.2,2) node{P2};
%     \end{tikzpicture}
%     \caption{Caption}
%     \label{fig:placeholder}
% \end{figure}

\subsection{Statistical convergence}

\begin{figure*}
    \centering
    \includegraphics[width=0.7\linewidth]{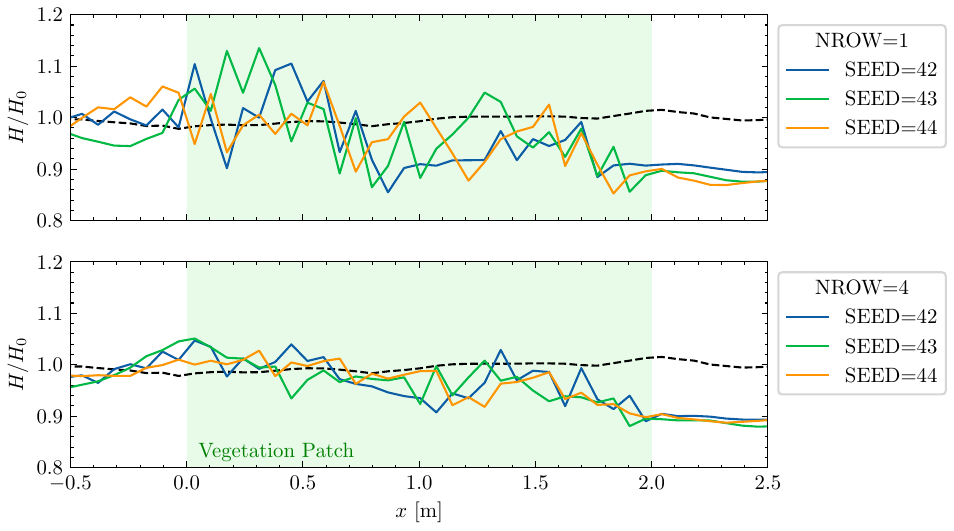}
    \caption{Normalised wave height along the channel. Green shaded region corresponds to the vegetated region (NCOL=30), and the black dashed line is the control wave height without vegetation.}
    \label{fig:sensitivity}
\end{figure*}

Because the cylinder heights are generated using a random number generator (RNG), in this subsection, we examine the sensitivity of the simulated wave attenuation to the RNG seed as well as the number of rows (channel width-wise) and columns (channel length-wise) of the cylinder array included in the simulation. The wave conditions considered are $T=1.26\,\mathrm{s}$, $H_0=0.1\,\mathrm{m}$, $h=0.6\,\mathrm{m}$, corresponding to $\beta h=1$ and $kh=1.64$.  Figure~\ref{fig:sensitivity} shows the spatial evolution of the simulated wave height along the channel, where the top panel is simulated with 1 rows and bottom panel is simulated with 4 rows. Increasing the number of cylinder rows reduces noise within the vegetation patch and leads to a more consistent transmitted wave height, indicating statistical convergence. 

\begin{figure*}
    \centering
    \includegraphics[width=0.73\linewidth]{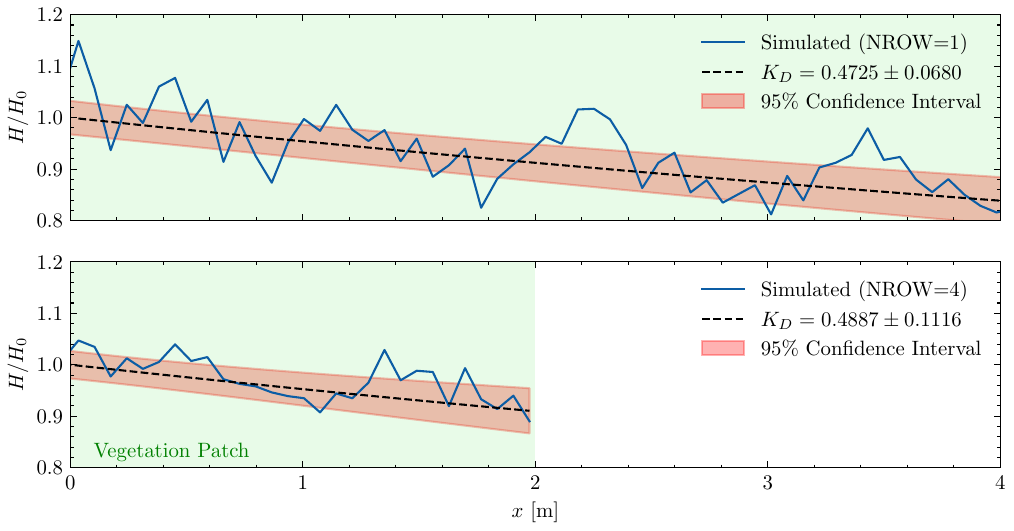}
    \caption{Fitted empirical wave decay relation (black dashed line) and the corresponding regression confidence interval (red band). The top panel corresponds to a configuration with one single rows and $L_\text{veg}=4\,\mathrm{m}$, whereas the bottom panel shows a case with four rows and $L_\text{veg}=2\,\mathrm{m}$.}
    \label{fig:fitted}
\end{figure*}

Finally, \eqref{eq:a} is fitted to the wave height measured within the vegetation patch to estimate the wave decay coefficient $K_D$. The fitted values of $K_D$ and their associated standard errors are shown in Figure~\ref{fig:fitted}. We note that doubling either the number of rows or number of columns both reduces the uncertainty in the estimated $K_D$. In practice, however, doubling the number of columns (i.e. longer vegetation patch) leads to a greater reduction in uncertainty than doubling the columns. Therefore, using a single row of cylinders with periodic side boundaries provides an efficient approach for simulating a patch of mangrove forest. 

\subsection{Calibration of $C_D$ with respect to $\beta h$}

To investigate the influence of the vegetation profile on the drag coefficient, we perform a series of numerical simulations for $\beta h=0$, $1$, and $2$ under the wave conditions listed in Table~\ref{tab:cases}. The water depth is fixed at $h=1.2\,\mathrm{m}$ for all cases. Based on the findings of the previous section, the vegetation field is represented by a single row of cylinders with periodic side boundaries. A total of $72$ rows are used, corresponding to a vegetation patch length of $L_{\mathrm{veg}}=5\,\mathrm{m}$.

\begin{figure}
    \centering
    \includegraphics[width=0.9\columnwidth]{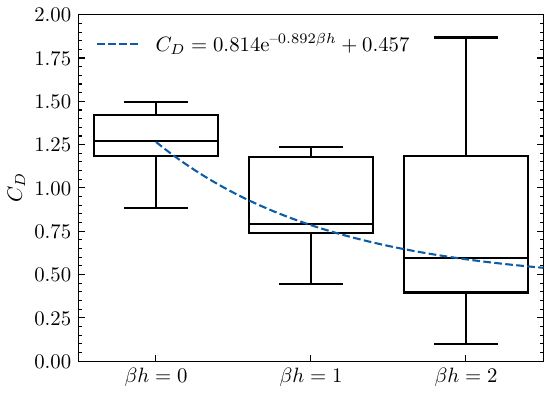}
    \caption{Box plot of the distribution of drag coefficient across various values of $\beta h$.}
    \label{fig:cd_box}
\end{figure}

The resulting distributions of $C_D$ are shown in Figure~\ref{fig:cd_box}. Some variability is observed across the different wave conditions, especially for larger $\beta h$. In general, smaller values of $\beta h$ correspond to larger drag coefficients. Physically, a smaller $\beta h$ indicates a greater concentration of vegetation frontal area near the water surface, where wave orbital velocities are typically strongest. As a result, the vegetation interacts more effectively with the wave-induced flow, leading to enhanced momentum extraction and a larger effective drag coefficient.

Conversely, increasing $\beta h$ shifts a larger fraction of the vegetation frontal area towards the lower water column, where wave orbital velocities are weaker. The vegetation therefore experiences reduced hydrodynamic forcing, resulting in a lower calibrated value of $C_D$. The observed trend is consistent with the wave attenuation results presented earlier, which showed that mangrove profiles with larger $\beta h$ are generally less effective at attenuating waves. 
% Despite the variation in $C_D$, Figure~\ref{fig:cd_box} indicates that the calibrated values remain within a relatively narrow range for each vegetation profile. This suggests that the exponential-profile representation captures the dominant influence of root morphology on wave attenuation and provides a robust basis for estimating drag coefficients across different mangrove configurations. 
The calibrated values also support the use of $\beta h$ as a compact descriptor linking mangrove morphology, hydrodynamic drag, and wave attenuation performance.

\begin{table}[htb]
    \centering
    \begin{tabular}{c c c}
        \hline
        Case & Wave Height $H_0 \,[\mathrm{m}]$ & Wave Period $T\,[\mathrm{s}]$ \\
        \hline
        1 & 0.1 & 1.003 \\
        2 & 0.1 & 1.430 \\
        3 & 0.1 & 2.197 \\
        4 & 0.1 & 3.871 \\
        5 & 0.2 & 1.430 \\
        6 & 0.2 & 2.197 \\
        7 & 0.2 & 3.871 \\
        \hline
    \end{tabular}
    \caption{List of wave conditions simulated. }
    \label{tab:cases}
\end{table}

\section{Conclusion}

% We introduced a practical parametrisation of the mangrove vegetation profile as a function of elevation, which can be readily fitted to field data and used to compare different mangrove species within a common framework. Using the mangrove profile, we formulated a wave attenuation formula and quantifies the wave attenuation efficiency based on wave regime, revealing that mangroves filter specific frequencies which depends on the root characteristic. Furthermore, we quantify the impact of sea-level rise on mangrove efficiency and demonstrate that in some specific cases, the wave attenuation capability of mangrove can increase under sea level rise. 

% Furthermore, we proposed a simplified mangrove representation based on vertical cylindrical elements that is suitable for laboratory experiments and numerical simulations, and can be configured to reproduce a prescribed vegetation drag profile through appropriate selection of cylinder height distribution, spacing, and diameter. Finally, we demonstrated that straight cylinder arrays are reasonable proxy for representing mangrove forest and that wave attenuation predictions are robust to the variability in root height. The framework established here forms the basis for subsequent parametric investigations and drag coefficient quantification. 

We introduced a practical parameterisation of mangrove vegetation profiles as a continuous function of elevation. The proposed representation can be readily calibrated from field measurements and provides a unified framework for quantifying species, site, and region specific mangrove characteristics. This enables local variations in root morphology to be systematically incorporated into wave attenuation assessments.

Using this vegetation profile, we derived an analytical wave attenuation model and a normalised wave decay coefficient that depends only on the dimensionless parameters $kh$ and $\beta h$. The model reveals that mangrove wave attenuation is frequency dependent and that different root morphologies preferentially attenuate different wave regimes. In particular, we identified a species-dependent critical wave regime at which wave attenuation is least effective. This behaviour is absent in conventional vertically uniform vegetation models and highlights the importance of accurately representing the vertical distribution of mangrove roots.

Building on this framework, we quantified the sensitivity of wave attenuation efficiency to sea-level rise through the dimensionless parameter $m$. The analysis shows that sea-level rise does not necessarily reduce attenuation efficiency; under certain wave and vegetation conditions, wave attenuation may in fact be enhanced. Furthermore, combining $m$ with IPCC AR6 sea-level-rise projections enables direct estimates of annual attenuation-efficiency loss under future climate scenarios. Beyond providing a quantitative measure of SLR vulnerability, the proposed framework offers a practical tool for identifying mangrove ecosystems that are most resilient to sea-level rise, thereby supporting evidence-based decision-making for nature-based solutions, ecosystem restoration, and long-term coastal adaptation planning.

Finally, we demonstrated that arrays of straight cylinders with exponentially distributed heights provide a reasonable proxy for mangrove root systems and for studying spatially inhomogeneous aquatic vegetation. The framework enables a prescribed vegetation drag profile to be reproduced through suitable choices of cylinder height, spacing, and diameter. Using this representation, we calibrated the drag coefficient $C_D$ across a range of $\beta h$ values, thereby establishing a direct link between vegetation morphology and wave attenuation performance. The resulting methodology provides a practical foundation for laboratory experiments, numerical modelling, and future investigations of vegetation-wave interactions.

\appendix
\section{Explicit expression for the SLR sensitivity parameter}
\label{sec:slr_appendix}

The SLR sensitivity parameter $m$ is defined as $m=h\eta'(0)$ where $\eta$ is the attenuation efficiency introduced in Section~\ref{sec:slr}. The prime denotes derivative with respect to $\Delta h$. The derivative is then evaluated at $\Delta h=0$ and multiply by a factor of $h$ such that $m$ is dimensionless and depends only on $kh$ and $\beta h$. The explicit expression for $m$ is given by 
\begin{subequations}
\begin{multline}
    m = 2 - \frac{\beta h}{\mathrm{e}^{\beta h}-1} - kh\frac{3\cosh(kh)\sinh(kh)+kh}{\sinh^2(kh)+kh\tanh(kh)}  \\
    % -\frac{\psi(kh,\beta h)}{\phi(kh,\beta h)}, 
    + \frac{\mathrm{e}^{\lambda_1} + 3\mathrm{e}^{\lambda_2} + 3\mathrm{e}^{\lambda_3} + \mathrm{e}^{\lambda_4}}{\frac{\mathrm{e}^{\lambda_1}-1}{\lambda_1} + 3\frac{\mathrm{e}^{\lambda_2}-1}{\lambda_2} 
    + 3\frac{\mathrm{e}^{\lambda_3}-1}{\lambda_3} + \frac{\mathrm{e}^{\lambda_4}-1}{\lambda_4}}.
\end{multline}
where 
\begin{align}
    \lambda_1 &= 3kh-\beta h, \\
    \lambda_2 &= kh-\beta h, \\
    \lambda_3 &= -kh-\beta h, \\
    \lambda_4 &= -3kh-\beta h.
\end{align}
\end{subequations}
% \begin{multline}
%     - \frac{T[\mathrm{e}^x](kh,\beta h)}{T[(\mathrm{e}^{x}-1)/x](kh,\beta h)}
%     T[f(x)](k,\beta) = f(3k-\beta) + 3f(k-\beta ) \\+ 3f(-k-\beta ) + f(-3k-\beta).
% \end{multline}
This expression is used to generate the results presented in Figure~\ref{fig:slr}. 

% \section{Derivation of the wave attenuation formula}

% Assuming zero current and a constant rate of wave energy dissipation, we have 
% \begin{equation} \label{eq:energy}
%     \frac{\partial}{\partial x}(Ec_g) = -\epsilon, \qquad E=\frac{1}{8}\rho g H(x)^2,
% \end{equation}
% where $E$ is the wave energy per unit area and $\epsilon$ is the dissipation rate due to the (time-averaged) drag force of the vegetation patch. % The vegetation resistance is given by the quadratic drag formula 
% % \begin{equation}
% %     f_D(z) = \frac{1}{2}\rho C_Da(z)u|u|.
% % \end{equation}
% % Here, $a(z)$ is the effective width of the vegetation per unit area. In practice, this can be measured by multiplying the number of roots of a mangrove tree at various elevations, by the mean root diameter, and the number of trees per unit area. 
% From Airy wave theory, the horizontal orbital velocity and the surface elevation are given by %$u = a\sigma\cos(\sigma t+\phi)\cosh(kz)/\sinh(kh)$
% \begin{align}
%     u(x,z,t) &= \frac{H}{2}\sigma \frac{\cosh(kz)}{\sinh(kh)}\cos(kx-\sigma t), \\ \eta(x,t) &= \frac{H}{2}\cos(kx-\sigma t). 
% \end{align}
% The energy dissipation rate is thus
% \begin{align}
%     \epsilon &= \frac{1}{T}\int_0^T \int_0^h f_D u\,\mathrm{d}z\,\mathrm{d}t \\
%     &= \frac{2}{3\pi}\rho C_D N_\text{tree}\left(\frac{H\sigma}{2\sinh(kh)}\right)^3\phi
% \end{align}
% Finally, substituting this into \eqref{eq:energy} and solving for $H(x)$ give the wave attenuation formula \eqref{eq:wave_decay}. 

\section*{Code Availability}
The source code for the numerical model developed in this study is open-source and available on GitHub at \url{https://github.com/pke1029/Openfoam-Inhomogeneous-Mangrove-Wave-Attenuation}. The repository contains the mesh generation tool as well as the OpenFOAM cases used to generate the simulation results presented in Section 3.

\section*{Acknowledgement}

This research/project is supported by the National Research Foundation, Singapore, and the National Parks Board, Singapore under its Marine Climate Change Science Programme (MCCS Award NRF-MCCS21-1-2-0001). Any opinions, findings and conclusions or recommendations expressed in this material are those of the author(s) and do not reflect the views of National Research Foundation, Singapore and National Parks Board, Singapore. 

\printcredits

\bibliographystyle{cas-model2-names}

\bibliography{references.bib}


% Author biography without author photo.
% Author biography. Author biography. Author biography.
% Author biography. Author biography. Author biography.
% Author biography. Author biography. Author biography.
% Author biography. Author biography. Author biography.
% Author biography. Author biography. Author biography.
% Author biography. Author biography. Author biography.
% Author biography. Author biography. Author biography.
% Author biography. Author biography. Author biography.
% Author biography. Author biography. Author biography.

% \bio{figs/cas-pic1}
% Author biography with author photo.
% Author biography. Author biography. Author biography.
% Author biography. Author biography. Author biography.
% Author biography. Author biography. Author biography.
% Author biography. Author biography. Author biography.
% Author biography. Author biography. Author biography.
% Author biography. Author biography. Author biography.
% Author biography. Author biography. Author biography.
% Author biography. Author biography. Author biography.
% Author biography. Author biography. Author biography.

% \bio{figs/cas-pic1}
% Author biography with author photo.
% Author biography. Author biography. Author biography.
% Author biography. Author biography. Author biography.
% Author biography. Author biography. Author biography.
% Author biography. Author biography. Author biography.

\end{document}